\documentclass[12pt]{article}
\usepackage{graphicx}
\usepackage{amsfonts}

\newcommand\eqref[1]{(\ref{#1})}

\begin{document}

\renewcommand\baselinestretch{1.6}\small\normalsize

\begin{center}
{\bf \Large {Differential geometry of density states}}\\
\end{center}

\begin{center}
{V.I. Man'ko,$^1$ G. Marmo,$^2$ E.C.G. Sudarshan$^3$ and F.
Zaccaria$^2$}
\end{center}

\begin{center}
$^1$P.N. Lebedev Physical Institute, Leninskii Prospect, 53,
Moscow 119991 Russia\newline $^2$Dipartimento di Scienze Fisiche,
Universit\`{a} ``Federico II'' di Napoli\\ and Istituto Nazionale
di Fisica Nucleare, Sezione di Napoli,\\ Complesso Universitario
di Monte Sant Angelo, Via Cintia, I-80126 Napoli, Italy\newline
$^3$Physics Department, Center for Particle Physics, University of
Texas, 78712 Austin, Texas, USA
\end{center}

\begin{center}
emails:~~{\small\textbf{manko@sci.lebedev.ru; marmo@na.infn.it;
sudarshan@physics.utexas.edu;\\zaccaria@na.infn.it }}
\end{center}

\begin{abstract}
We consider a geometrization, i.e., we identify geometrical
structures, for the space of density states of a quantum system.
We also provide few comments on a possible application of this
geometrization for composite systems.
\end{abstract}

\textbf{keywords}: {quantum states, geometric structures,
Fubini--Study projective geometry. }

\section{Introduction}

The notions of "states" and "observables" are fundamental in quantum
mechanics. The space of states is usually assumed to be a vector space due
to the introduction by Dirac~\cite{Dirac} of the superposition principle as
a fundamental principle of quantum theory. It is true however that one may
consider superposition rules of solutions also for nonlinear evolution
equations~\cite{Carinena}, this would give rise to a dynamical version of
superposition (i.e., all superpositions of solutions for a given evolution
equation represent new solutions).

The interpretation of wave functions as probability amplitudes suggested
that the state space be identified as a Hilbert space, i.e., a vector space
with an inner product.

The superposition principle, when applied to product states of composite
systems, gives rise to the fundamental aspects of quantum nonseparability.
In fact, Schr\"odinger~\cite{Schroedinger} had identified entanglement as a
characteristic ingredient of quantum multipartite systems; superposition
appears in classical wave phenomena and we have water waves and sound waves,
but quantum entanglement is totally new.

While this entanglement has many observed consequences, it becomes
spectacularly manifest in the violation of Bell's inequality. A particular
use of it was the construction of "coherent state" of photons to represent
electromagnetic fields.

Usually the implementation of the superposition principle at the level of
solutions of evolution equations requires these equations to be linear. Some
limitations arise from superselection rules but this will not concern us
here.

In spite of this fundamental principle, usually one identifies states with
rays generated by Hilbert space vectors rather than with the vectors
themselves. This identification requires some additional ingredient to be
able to discuss interference phenomena within the framework of rays. This
becomes more evident if we use the identification of states as given by
rank-one projectors, indeed, the usual superposition rule of two of these
projectors will give, in general, as a result an operator of rank two. A way
to handle this problem has been proposed recently~\cite{MMSZ}, see also in
this connection the recent paper~\cite{Gorelli}. A net result of the trading
of vectors with equivalence classes is that we have a carrier space for
quantum evolution which is not a linear space any more but a differential
manifold (the complex projective space).

As for "observables", they are usually associated with measure operations.
Measurements are described by means of Hermitian operators and often
introduced as additional independent ingredients for the description of
quantum phenomena. Even though Hermitian operators do not constitute an
associative algebra, one usually considers them as part of the algebra of
operators acting on the Hilbert space of quantum vectors. In the early
times~ \cite{Jordan} of quantum mechanics, to have a binary product within
the space of Hermitian operators. the Jordan product was introduced; it is a
commutative product but not associative. By mapping Hermitian operators into
the Lie algebra of the unitary group we get a binary product, the commutator
or Lie product, which is a bilinear inner composition rule. These two
products are enough to capture the essential ingredients of the measurement
rules.

From what we have said about the identification of physical states
with
points of the complex projective space (associated with the Hilbert space $%
\mathcal{H}$), it is clear that the manifold structure of this space
requires that we replace all objects, whose definition depends on the
existing linear structure on $\mathcal{H}$, with "tensorial objects", i.e.,
geometrical entities which preserve their meaning under general
transformations and not just linear ones. This "tensorial" viewpoint has
been encoded into the differential-geometric approach to quantum mechanics
which has been undertaken by a large number of physicists~\cite{all}. For a
recent textbook treatment, we refer to the nice book by Chruscinski and
Jamiolkowski~\cite{Ch-Jam}.

Bearing in mind these last remarks, in this paper we shall consider the
Hilbert space $\mathcal{H}$ as a real differential manifold with additional
structures carrying an action of the unitary group. To let the geometrical
structures to emerge neatly without the technicalities due to the infinite
dimension we shall restrict ourselves to finite dimensional Hilbert spaces.

Therefore, the differential-geometric point of view is implemented
by considering our relevant spaces as real differential manifolds.
The complex structure of the standard Hilbert space is considered
to be an additional structure on the real differential manifold.

This paper is addressed to theoretical physicists at home with the
geometrical structures employed within the geometrical approach to classical
mechanics.

\section{The space of state vectors as a differential manifold}

Starting with the complex Hilbert space $\mathcal{H}$, to deal with its real
differential structure, we consider its ''realification'' $\widetilde{%
\mathcal{H}}$ with the additional structures arising from the Hermitian
inner product, i.e., the real part which defines a positive definite
Riemannian structure (Euclidean product) $g$, the imaginary part defining a
not degenerate skew symmetric bilinear product which is a symplectic
structure $\omega $ and a map connecting the two which corresponds to the
multiplication by the imaginary unit, the complex structure $J$ satisfying
the properties $J^{2}=-\mathbf{1}$ and $g(x,y)=\omega (Jx,y)$.

These three structures join together to define the Hermitian inner product
\begin{equation}
h(x,y)=g(x,y)+i\omega (x,y).  \label{1}
\end{equation}
To turn these entities into tensors, we consider $x,y$ as vector fields on
the manifold $\widetilde{\mathcal{H}}$ while eq.~(\ref{1}) is thought of as
the evaluation on vectors in $T_{\psi }\widetilde{\mathcal{H}}$ at the point
$\psi \in \widetilde{\mathcal{H}}$. More specifically vector fields, $X:%
\widetilde{\mathcal{H}}\rightarrow T\widetilde{\mathcal{H}}\Leftrightarrow
\widetilde{\mathcal{H}}\times \widetilde{\mathcal{H}}$, and 1-forms $\alpha :%
\widetilde{\mathcal{H}}\rightarrow T^{\ast }\widetilde{\mathcal{H}}%
\Leftrightarrow \widetilde{\mathcal{H}}\times \widetilde{\mathcal{H}}^{\ast
}\Leftrightarrow \widetilde{\mathcal{H}}\times \widetilde{\mathcal{H}}$ will
be identified with their second component. $\widetilde{\mathcal{H}}$ and $%
\widetilde{\mathcal{H}}^{\ast }$ are identified by means of the Euclidean
inner product, associated with $g$ .

By using collective coordinates $\{x^{j}\}$, we would have
\begin{eqnarray}
&&g=g_{jk}dx^{j}\otimes dx^{k},  \nonumber  \label{2} \\
&&\omega =\omega _{jk}dx^{j}\wedge dx^{k}, \\
&&J=g^{jk}\omega _{kl}dx^{l}\otimes \frac{\partial }{\partial x^{j}}%
\,,\qquad j,k=\{1,2,\ldots ,2n\},  \nonumber
\end{eqnarray}
with the property $J^{2}=-\mathbf{1}$, \ and \
$g^{jk}g_{kl}=\delta _{l}^{j}$.

It is not difficult to show that if formula~(\ref{1}) defines a Hermitian
product we have
\begin{eqnarray}
&&(\mathrm{a)\qquad g(Jx,Jy)=g(x,y),\quad \forall\,\,x,y,\quad
g(Jx,y)+g(x,Jy)=0,}  \label{3a} \\
&&(\mathrm{b)\qquad\omega(Jx,Jy)=\omega(x,y),\qquad
\omega(Jx,y)+\omega(x,Jy)=0,}  \label{3b}
\end{eqnarray}
i.e., $J$ generates both finite and infinitesimal transformations which are
orthogonal and symplectic.

The vector space structure of $\mathcal{H}$ is associated with the dilation
vector field $\Delta$ given by
\begin{equation}  \label{4}
\Delta=x^j\,\frac{\partial}{\partial x^j}
\end{equation}
which is also known as Liouville vector field or the Euler operator.

By using the $(1-1)$-tensor field $J$ we may define another vector field $%
\Gamma =J(\Delta )$. These two vector fields commute and generate a
foliation of $\ \widetilde{\mathcal{H}}-\{0\}$ in terms of two-dimensional
real vector spaces (strictly speaking, leaves are diffeomorphic with $%
\mathcal{R}^{2}-\{0\})$.

By means of $\Delta$ one defines homogeneous polynomial functions of degree $%
k$ by requiring $\Delta\cdot f=kf$. This definition has the advantage of
being coordinate independent, i.e., we may even perform nonlinear
transformations of coordinates.

The symplectic structure $\omega $ defines a Poisson tensor
\begin{equation}
\Lambda =\Lambda ^{jk}\frac{\partial }{\partial x^{j}}\wedge \frac{\partial
}{\partial x^{k}}  \label{5}
\end{equation}
with
\begin{equation}
\Lambda ^{jk}\omega _{kl}=\delta _{l}^{j},\qquad \Lambda ^{jk}=-\Lambda
^{kj}.  \label{6}
\end{equation}
It is also possible to consider the inverse of $g$, namely,
\begin{equation}
G=G^{jk}\frac{\partial }{\partial x^{j}}\otimes \frac{\partial }{\partial
x^{k}}  \label{7}
\end{equation}
with
\begin{equation}
G^{jk}=G^{kj}\qquad \mbox{and}\qquad G^{jk}g_{kl}=\delta _{l}^{j}.  \label{8}
\end{equation}
They are related by $G=J\cdot \Lambda $. These tensors $\Lambda $ and $G$
allow us to define a Poisson bracket and a Riemann--Jordan bracket \cite
{Pizzocdeura} on smooth functions in $\mathcal{F}(\widetilde{\mathcal{H}})$
by setting, respectively,
\begin{equation}
\{f,g\}=\Lambda (df,dg)=\Lambda ^{kj}\left( \frac{\partial f}{\partial x^{j}}%
\frac{\partial g}{\partial x^{k}}\right)  \label{9}
\end{equation}
and
\begin{equation}
(f,g)=G(df,dg)=G^{kj}\left( \frac{\partial f}{\partial x^{j}}\frac{\partial g%
}{\partial x^{k}}\right) .  \label{10}
\end{equation}
By using $\Lambda $, we may consider the group of transformations which
preserve both $\Lambda $ and $\Delta $, we get in this way the group of real
linear symplectic transformations. By replacing $\Lambda $ with $G$, we
define the group of real linear orthogonal transformations. The intersection
of these two invariance groups defines the group of unitary transformations,
denoted by $U(n)$ when $\widetilde{\mathcal{H}}$ is assumed to be of (real)
dimensions $2n$.

Symmetric tensor fields $t=t_{kj}dx^{j}\otimes dx^{k}$ are converted into
quadratic functions $f_{t}$ by considering
\begin{equation}
2f_{t}=t(\Delta ,\Delta )=t_{jk}x^{j}x^{k},  \label{11}
\end{equation}
and similarly for higher order tensors. For skew-symmetric
2-tensors, we may define for $\gamma =\gamma _{jk}\,dx^{j}\wedge
dx^{k}$, $2f_{\gamma }=\gamma (\Delta ,J(\Delta ))$. (The factor
$2$ is very convenient if at some point we want to identify our
function with the energy, when it is the case.) When the
skew-symmetric 2-tensor coincides with the symplectic structure,
the corresponding function is the Hamiltonian function generating
the one-parameter group of unitary transformations which consists
of multiplication by a phase.

Any linear operator $A:\widetilde{\mathcal{H}}\rightarrow \widetilde{%
\mathcal{H}}$ can be converted into a $(1-1)$ tensor field by setting
\begin{equation}
T_{A}=A_{k}^{j}\,dx^{k}\otimes \frac{\partial }{\partial x^{j}}  \label{12}
\end{equation}
or into a vector field
\begin{equation}
X_{A}=T_{A}(\Delta )=A_{k}^{j}x^{k}\frac{\partial }{\partial x^{j}}\,.
\label{13}
\end{equation}
along with \ $Y_{A}=T_{A}(J(\Delta ))$ . For $\ A$ hermitian operator, $%
X_{A} $ corresponds to the gradient vector field, while $Y_{A}$
corresponds to the Hamiltonian vector field associated with the
expectation value of the operator.

The association of $A$ with $T_{A}$ is an associative algebra isomorphism,
while the association of $A$ with $X_{A}$ allows one to capture only the Lie
algebra structure.

By using $g$ and $\omega $, it is possible to associate a complex valued
quadratic function on $\widetilde{\mathcal{H}}$ with any linear
transformation $A$ by setting
\begin{equation}
2f_{A}=g\Big(\Delta ,T_{A}(\Delta )\Big)+i\omega \Big(\Delta ,T_{A}(\Delta )%
\Big)=g_{jk}A_{l}^{k}x^{j}x^{l}+i\omega _{jk}A_{l}^{k}x^{j}x^{l},  \label{14}
\end{equation}
equivalently, on $\mathcal{H}$, we could write $2f_{A}(\psi )=\langle \psi
\mid A\psi \rangle $.

All our constructions have been written in an implicit form to
exhibit their independence of the choosen coordinates and to hint
at the fact that they remain true at the level of infinite
dimensional Hilbert spaces, whenever the relative tensors are
defined.

Because we shall be mainly interested in the ''realification'' of operations
taking place on the complex Hilbert space, we shall always consider $(1-1)$%
-tensor fields $T_{A}$ associated with complex-linear operators, i.e., $%
T_{A}\cdot J=J\cdot T_{A}$

This amounts to consider only complex linear transformations, i.e. only real
representations of $GL(n,\Bbb{C)}$ .

It is now a simple result following from computations that for complex
valued functions
\begin{equation}
\{f_{A},f_{B}\}=-if_{[A,B]}  \label{15}
\end{equation}
and
\begin{equation}
(f_{A},f_{B})=f_{(AB+BA)},  \label{16}
\end{equation}
i.e., for quadratic functions associated with complex linear operators, we
recover the Lie product and the Jordan product by using the Poisson tensor $%
\Lambda $ and the Riemannian tensor $G$, respectively.

We have now the possibility of characterizing canonical
transformations in a way that turns out to be useful when dealing
with quantum gates. We recall that canonical transformations are
implicitly defined by the property of leaving the Poisson brackets
invariant. However, when dealing with a transformation from
$(q,p)$ to $(Q,P)$ variables, a different characterization is
obtained by requiring that
\[
p^{a}dq_{a}-P^{b}dQ_{b}=dS(q,Q).
\]
In the extended formalism \cite{EMS} this would be
\[
(p^{a}dq_{a}-H\,dt)-(P^{b}dQ^{b}-K\,dt)=dW(q,Q;t).
\]
If \ we want to make contact with standard Hilbert space approach, we may
use complex coordinates, say $\psi =q+ip$, $\phi =Q+iP$, \ and we may write
previous equations in the form
\[
\frac{1}{2}\,i\left[ \phi ^{k}d\phi _{k}^{\ast }+\phi _{k}^{\ast }d\phi
^{k}-2d\phi ^{k}\phi _{k}^{\ast }-\left( \psi ^{k}d\psi _{k}^{\ast }+d\psi
^{k}\psi _{k}^{\ast }\right) +2d\psi ^{k}\psi _{k}^{\ast }\right] =dS(\psi
^{k},\phi ^{k}),
\]
or directly
\[
i(d\psi ^{k}\psi _{k}^{\ast }-d\phi ^{k}\phi _{k}^{\ast })=dS(\phi ^{k},\psi
^{k}).
\]
This equation may be spelled out into
\[
i\psi _{k}^{\ast }=\frac{\partial S}{\partial \psi ^{k}},\qquad -i\phi
_{k}^{\ast }=\frac{\partial S}{\partial \phi ^{k}}.
\]

\textbf{Remark}:$\quad $ By making different choices of the independent
variables, say instead of $\psi ,\phi $ we could use $(\psi ,\phi ^{\ast })$%
, $(\psi ^{\ast },\phi ^{\ast })$, $(\psi ^{\ast },\phi )$, we
would have different expressions for the generating functions
\cite{EMS}. If we require the transformation to be linear, we
start with a quadratic generating functions $S$. In general, as
functions of complex variables, they should be at least analytic
in the relevant variables. By requiring that the resulting
transformation preserves $J$ we find that eventually our
transformations will be unitary. If $\phi ^{k}=U_{j}^{k}\psi _{j}$
is a unitary transformation connecting two different bases,one
obtains as a generating function $S=i\phi _{k}^{\ast
}U_{j}^{k}\psi ^{j}$. However, one should bear in mind that the
class of canonical transformations is much more larger than the
class of unitary transformations, and moreover they will be well
defined also on the space of rays, the complex projective space.

In the simplest case of $\mathcal{H}=\mathcal{C}^{2}$, we may write few
generating functions which describe some well known quantum gates.

For beam splitter or Hadamard gate, we have
\[
S_{\mathrm{H}}=\frac{i}{\sqrt{2}}\left( \phi _{1}^{\ast }\psi ^{1}+\phi
_{2}^{\ast }\psi ^{1}+\phi _{1}^{\ast }\psi ^{2}-\phi _{2}^{\ast }\psi
^{2}\right) ,
\]
similarly for the phase gate $S=i(\phi _{1}^{\ast }\psi ^{1}-\phi _{2}^{\ast
}\psi ^{2})$, and for the phase-shift we have $i\left( \phi _{1}^{\ast }\psi
^{1}+e^{i\theta }\phi _{2}^{\ast }\psi ^{2}\right) $.

As it is well known, generating functions add to represent the composition
of transformations. Having represented gates in terms of generating
functions, we are no more restricted to linear spaces.

From now onwards, whenever there is no danger of confusion we
shall \ also make use, in the algebraic or linear setting, of
complex coordinates so that the comparison with standard quantum
mechanics becomes more transparent. While,whenever we deal with
non-linear transformations or tensorial aspects \ of quantum
mechanics or differential geometrical aspects, our treatment only
considers real differential structures.

\section{\protect\bigskip\ The unitary group as a group of canonical
transformations: momentum map}

The group of linear transformations preserving the triple $(g,\omega ,J)$ is
the unitary group denoted by $U(n)$, the associated Lie algebra will be
denoted $u(n)$. If we denote by $\mathcal{X(\widetilde{\mathcal{H}})}$ the
Lie algebra of vector fields on $\widetilde{\mathcal{H}}$, we have a Lie
algebra homomorphism $u(n)\rightarrow \mathcal{X(\widetilde{\mathcal{H}})}$.
With any Hermitian operator $A$ we may associate a vector field
\begin{equation}
X_{A}=-\frac{i}{\hbar }\,T_{A}(\Delta ),  \label{17}
\end{equation}
which is the infinitesimal generator of the one-parameter group of unitary
transformations
\begin{equation}
U(\alpha )=e^{-i\alpha A/\hbar },  \label{18}
\end{equation}
where the parameter $\alpha $ is such that the product of the physical
dimensions of $\alpha $ and $A$ has the dimension of an action. With a
slight abuse of notation, denoting by $iu(n)$ the set of Hermitian
operators, we have a map linear in the second argument
\begin{equation}
F:\widetilde{\mathcal{H}}\times iu(n)\rightarrow \mathcal{R}  \label{19}
\end{equation}
specifically
\begin{equation}
2F(\psi ,A)=\langle \psi \mid A\psi \rangle .  \label{20}
\end{equation}
It has the property
\begin{equation}
\{F(A),F(B)\}=iF([A,B]),  \label{21}
\end{equation}
where $F(A):\widetilde{\mathcal{H}}\rightarrow {\mathcal{R}}$ is defined out
of $F$ in an obvious way. By using the Cartesian property of maps, $\mathcal{%
F}\Big(\widetilde{\mathcal{H}}\times U(n)\Big)=\mathcal{F}\Big(\widetilde{%
\mathcal{H}},\mathrm{Lin}(U(n),\mathcal{R})\Big)$, we may define also
\begin{equation}
\hat{F}:\widetilde{\mathcal{H}}\rightarrow u^{\ast }(n)=\mathrm{Lin}(u(n),%
\mathcal{R}).  \label{22}
\end{equation}
This map is usually called the momentum map associated with the
symplectic action of the group $U(n)$ on
$\widetilde{\mathcal{H}}$.

It is useful to present explicitly the momentum map associated
with $U(n)$ acting on $\widetilde{\mathcal{H}}$, i.e., the complex
Hilbert space, in terms of Dirac notation and using the
identification of $u(n)$ with its dual by means of the
Cartan--Killing metric structure on $u(n)$ we have
\begin{equation}  \label{23}
\hat F(\psi)=-i\mid\psi\rangle\langle\psi\mid.
\end{equation}

We find the remarkable result that the unit sphere of the Hilbert space can
be imbedded into $u^*(n)$ equivariantly with respect to the coadjoint action.

As a matter of fact, it is now possible to foliate $\widetilde{\mathcal{H}}%
-\{0\}$ with the involutive distributions of the vector fields $\Delta $ and
$J(\Delta )$, to obtain as quotient a real differential manifold,
diffeomorphic with the complex projective Hilbert space $\mathcal{P}(%
\widetilde{\mathcal{H}})$. It is not difficult to show that $J(\Delta )$ is
the Hamiltonian vector field associated with the function $f_{\mathbf{1}}=%
\frac{1}{2}\langle \psi \mid \psi \rangle $.

If we replace the evaluation function $2f_{A}(\psi )=\langle \psi \mid A\psi
\rangle $ with the expectation value of $A$ at $\psi $, say
\begin{equation}
\tilde{f}_{A}(\psi )=\frac{\langle \psi \mid A\psi \rangle }{\langle \psi
\mid \psi \rangle }\,,  \label{24}
\end{equation}
we find that $\tilde{f}_{A}(\psi )$ is invariant under $\Delta $ and $%
J(\Delta )$. The invariance under $\Delta $ is obvious. For the invariance
under $J(\Delta )$, we use the fact that $J(\Delta )$ is the Hamiltonian
vector field associated with $f_{\mathbf{1}}$ and $\{f_{\mathbf{1}%
},f_{A}\}=0 $ from (\ref{21}). Thus the algebra of functions defined by the
expectation values of Hermitian operators projects onto the quotient space
of $\widetilde{\mathcal{H}}-\{0\}$ with respect to the foliation defined by $%
\Delta $ and $J(\Delta )$. This algebra separates the points of the quotient
and therefore completely determines it.

\textbf{Remark. }Equivalently to obtain a differential manifold
diffeomorphic \ with \ the quotient complex projective space we consider the
Poisson bracket on $\widetilde{\mathcal{H}}$ and notice that the centralizer
of $\ 2f_{\mathbf{1}}=\langle \psi \mid \psi \rangle $ in $\mathcal{F}(%
\widetilde{\mathcal{H}},\mathcal{R}))$ is a Poisson subalgebra. The Poisson
bracket can be extended to complex valued function and contains the set of
quadratic complex valued functions associated with complex linear operators $%
A:{\mathcal{H}}\rightarrow {\mathcal{H}}$, i.e., $\tilde{f}_{A}(\psi )=\frac{%
\langle \psi \mid A\psi \rangle }{\langle \psi \mid \psi \rangle }$ . This
Poisson algebra generates the Poisson algebra of complex valued functions
defined on the complex projective space. Indeed, each function $\tilde{f}%
_{A}(\psi )$ is invariant under the infinitesimal action of $\Delta $ and $%
J(\Delta )$. The Poisson bracket on the complex projective space defines a
symplectic structure. The complex structure $J$ on $\widetilde{\mathcal{H}}$
induces a complex structure on ${\mathcal{P}}\widetilde{\mathcal{H}}$ by
setting $\hat{F}_{\ast }(\tilde{J}\,df):=J\Big(\hat{F}_{\ast }(df)\Big)$ for
any function on ${\mathcal{P}}\widetilde{\mathcal{H}}$. The invariance of $J$
under the action of $\Delta $ and $J(\Delta )$ shows that $J\Big(\hat{F}%
_{\ast }(df)\Big)$ is the pull-back of a 1-form on ${\mathcal{P}}\widetilde{%
\mathcal{H}}$ and therefore the left-hand side (i.e., $\tilde{J})$ is well
defined. Out of the symplectic structure and $\tilde{J}$ on ${\mathcal{P}}%
\widetilde{\mathcal{H}}$ we can construct a K\"{a}hler structure.

The projection $\tilde F:\widetilde{\mathcal{H}}-\{0\}\rightarrow{\mathcal{P}%
}\widetilde{\mathcal{H}}$, written in explicit form by means of the momentum
map associated with unitary transformations, and the required invariance
under $\Delta$ and $J(\Delta)$, has the form
\[
\tilde f(\psi)=-i\, \frac{\mid \psi\rangle\langle\psi\mid}{%
\langle\psi\mid\psi\rangle}\,.
\]
A connection 1-form for this projection can be given by requiring that $%
\Delta$ and $J(\Delta)$ are fundamental vector fields.

The connection 1-form can be written in compact form by using Dirac notation
\[
\theta =\frac{\langle \psi \mid d\psi \rangle }{\langle \psi \mid \psi
\rangle }\,.
\]
It is now sufficient to show that
\[
\theta (J(\Delta ))=i,\qquad \theta (\Delta )=1.
\]
By using real coordinates, say, $\psi =(q+ip)$, we find
\[
\frac{\psi ^{\ast }d\psi}{\langle \psi \mid \psi \rangle }=\frac{%
(q-ip)d(q+ip)}{q^{2}+p^{2}}=\frac{qdq+pdp}{q^{2}+p^{2}}+i\,\frac{qdp-pdq}{%
q^{2}+p^{2}}\,,
\]
while
\[
\Delta =q\,\frac{\partial }{\partial q}+p\,\frac{\partial }{\partial p}%
,\qquad J(\Delta )=q\,\frac{\partial }{\partial p}-p\,\frac{\partial }{%
\partial q}\,.
\]
Horizontal vectors are those vectors in $T_{\psi }\mathcal{H}$ which are in
the kernel of $\theta $. Usually one avoids dealing with additional terms
due to $\Delta $ by restricting very soon all considerations to the unit
sphere $S(\mathcal{H})=\left\{ \psi \in \mathcal{H},\langle \psi \mid \psi
\rangle =1\right\} .$

Our choice is dictated by the desire to keep separate notions which depend
on the chosen Hermitian structure from those which do not rely on it.

It is now possible to write an Hermitian \ tensor \ which is a
metric on horizontal vectors or equivalently on
${\mathcal{P}}{\mathcal{H}}$ (where \ \ it becomes the well-known
Fubini--Study metric tensor\cite{Study}) by setting
\[
\frac{(\psi \wedge d\psi )^{2}}{\langle \psi \mid \psi \rangle ^{2}}=\left\{
\frac{\langle d\psi \mid d\psi \rangle }{\langle \psi \mid \psi \rangle }-%
\frac{\langle \psi \mid d\psi \rangle \langle d\psi \mid \psi \rangle }{%
\langle \psi \mid \psi \rangle ^{2}}\right\} .
\]
(The use of vector-valued differential forms may be very convenient for
quick algebraic manipulations, some aspects are dealt with in\cite{Grab})

The real part of this Hermitian tensor will be the Riemannian
metric while the imaginary part will be the symplectic structure
on ${\mathcal{P}}\widetilde{\mathcal{H}}$.

The trick of restricting everything to the unit sphere $S({\mathcal{H}}%
)\subset {\mathcal{H}}$ hides the fact that the pull-back of the
Fubini--Study metric on ${\mathcal{P}}\widetilde{\mathcal{H}}$ is only
conformally related to the Euclidean metric on $\widetilde{\mathcal{H}}$.

\section{Density states}

Our imbedding of the complex projective space into $u^{\ast }(n)$ is
achieved by setting, for any equivalence class $[\psi ]$,
\begin{equation}
\lbrack \psi ]\rightarrow -i\,\frac{\mid \psi \rangle \langle \psi \mid }{%
\langle \psi \mid \psi \rangle }\,,  \label{25}
\end{equation}
which clearly does not depend on the representative choosen within $[\psi ]$.

By introducing the rank-one projector $\rho _{\psi }$,
\begin{equation}
\rho _{\psi }=\frac{\mid \psi \rangle \langle \psi \mid }{\langle \psi \mid
\psi \rangle }\,,  \label{26}
\end{equation}
we can write in standard notation
\begin{equation}
f_{A}(\psi )=\mbox{Tr}\,\rho _{\psi }A=Tr\left[ \left[ \psi \right] \left(
iA\right) \right] .  \label{27}
\end{equation}

Let us summarize the situation.

Out of the action of the unitary group on $\widetilde{\mathcal{H}}$, we have
constructed an imbedding of the complex projective space into the dual of
the Lie algebra of the unitary group, i.e., by means of the momentum map.
The target space of this map is a linear space, therefore, we have imbedded
our nonlinear differential manifold into a linear space, allowing us to
consider linear combinations of points in the image and therefore to go
beyond rays, or pure states. The equivariance of the momentum map means that
any unitary evolution on the Hilbert space (associated with the
Schr\"{o}dinger equation) will give rise to a unitary evolution on $u^{\ast
}(n)$ (associated with the von Neumann equation) \cite{Russ}.

The interpretation of $f_A([\psi])$ as an expectation value allows us to
consider probability distributions or averaging $A$ on a set of states with
appropriate weights. We may thus consider the average
\begin{equation}  \label{28}
\sum_kp_k\mbox{Tr}\,\rho_k A=\mbox{Tr}\,\rho A
\end{equation}
with
\[
p_k\geq 0\,\,\forall
k,\,\,\sum_kp_k=1\,\,\,\rho_k=\rho_k^\dagger,\,\,\rho_k^2=\rho_k,\,\, %
\mbox{tr}\,\rho_k =1.
\]
Out of points $(\rho_1,\rho_2,\ldots,\rho_k,\ldots)$ in the image of the
complex projective space, we have formed $\sum_kp_k\rho_k$ which can be
identified with an element in $u^*(n)$. The probabilistic interpretation
requires that we deal only with convex combinations.

Clearly we can extend our original Poisson bracket on expectation values
from $f_{A}(\psi )$ to $f_{A}(\rho )=\sum_{k}p_{k}f_{A}(\psi _{k})$ simply
by linearity
\begin{equation}
\{f_{A},f_{B}\}(\rho ):=\sum_{k}p_{k}f_{i[A,B]}(\psi
_{k})=\sum_{k}p_{k}\{f_{A},f_{B}\}(\psi _{k}).  \label{29}
\end{equation}
Similarly, the Riemann--Jordan bracket can be extended by linearity
\begin{equation}
(f_{A},f_{B})(\rho ):=\sum_{k}p_{k}(f_{A},f_{B})(\psi _{k}).  \label{30}
\end{equation}
From now onwards we shall denote all convex combinations of pure states as $%
\mathcal{D(\mathcal{H})}$ and call them density states.

\section{Geometrical structures on the space of density states}

We shall now consider the mathematical structures available on $\mathcal{D(%
\mathcal{H})}$. When we want to use differential-geometric properties of $%
\mathcal{D(\mathcal{H})}$, we shall always consider it as a real
differential manifold with boundary and denote it as $\mathcal{D(\widetilde{%
\mathcal{H}})}$ imbedded into the real vector space $u^{\ast }(n)$.

On $\mathcal{D}(\widetilde{\mathcal{H}})$ there is a Poisson tensor $%
\widetilde\Lambda$ associated with brackets (\ref{29}) and a metric tensor $%
\widetilde G$ associated with bracket (\ref{30}). Obviously $%
\widetilde\Lambda$ is degenerate, the kernel being associated with
Casimir functions. If we use expectation values $f_A(\rho)$, we
can define a partial complex structure by setting \cite{Aldaya}
\begin{equation}  \label{31}
J\widetilde G(df_A):=\widetilde\Lambda(df_A)
\end{equation}
showing that gradient-vector fields associated with Casimir functions must
be in the kernel of $J$. Therefore $J^2=-\mathbf{1}$ only when we restrict
it to combinations of Hamiltonian vector fields. We also notice that
Hamiltonian vector fields are always tangent to the topological boundary of $%
\mathcal{D}(\widetilde{\mathcal{H}})$.

In summary, we may say that the tangent space of
$\mathcal{D}(\widetilde {\mathcal{H}})$, in its internal points,
is spanned by the Hamiltonian vector fields and the gradient
vector fields associated with Casimir functions.

As $\mathcal{D}(\widetilde{\mathcal{H}})$ is the union of
symplectic orbits of the coadjoint action of $U(n)$ on $u^{\ast
}(n)$, on each orbit there is a symplectic structure defined by
projecting from the group onto the orbit \cite{Mukunda}
\begin{equation}
\omega _{\rho }=d\,\mbox{Tr}\left( \rho U^{\dagger }dU\right) =-\mbox{Tr}%
\left( \rho U^{\dagger }dU\wedge U^{\dagger }dU\right) .  \label{32}
\end{equation}
The boundary is a stratified manifold,being the union of
symplectic orbits of different dimensions passing through density
matrices of not maximal rank.It should be remarked that while the
two-form is well defined on the orbit, the one-form $U^{\dagger
}dU$ is a Lie algebra valued one-form on the unitary group $U(n)$
which does not descent to the orbit. It is not difficult to show
that the kernel of $\omega _{\rho }$, Ker$\,\omega _{\rho } $, is
spanned by infinitesimal generators of the isotropy group (or
stability group)of $\rho $ under the coadjoint action of the
unitary group.

Previous considerations show that
$\mathcal{D}(\widetilde{\mathcal{H}})$ has an inverse image under
the momentum map $\mu:T^*U(n)\rightarrow u^*(n)$ . We may consider
$\mu^{-1}\Big(\mathcal{D}(\widetilde{\mathcal{H}})\Big)$ and use
the geometric structures available on $T^*U(n)$ which is
diffeomorphic with a subgroup of $GL(n,\mathcal{C})$. In
particular, $T^*SU(n)$ is symplectomorphic with the group
$SL(n,\mathcal{C})$ considered as a Drinfeld double \cite{Marmo}.

In particular, any torus action coming from a Cartan subalgebra in
$U(n)$ gives rise to a complexified torus action on $T^{\ast
}U(n)$, no more unitary. This violation of unitarity seems to have
an interpretation in terms of quantum measurements and provides a
possible useful description of the wave function collapse
\cite{Benvegnu}. We shall develop these considerations elsewhere.

\section{Example: a two-level system}

In this section, we shall describe our previous constructions in terms of $%
\mathcal{H}=\mathcal{C}^2\equiv\mathcal{R}^4$, i.e., a two-level quantum
system.

We introduce an orthogonal basis $\mid e_{1}\rangle $ and $\mid e_{2}\rangle
$ and any vector $\mid \psi \rangle $ will be decomposed into $\mid \psi
\rangle =z_{1}\mid e_{1}\rangle +z_{2}\mid e_{2}\rangle $, therefore $%
\mathcal{C}^{2}$ will be parametrizied by complex coordinates $(z_{1},z_{2})$%
. To deal with its realification $\mathcal{R}^{4}$, we consider also real
coordinates defined by the formulae $z_{1}=q_{1}+ip_{1}$, $%
z_{2}=q_{2}+ip_{2} $.

The unitary group $U(2)$ is realized by requiring that it preserves the
quadratic function $z_{1}^{\ast }z_{1}+z_{2}^{\ast }z_{2}$, or equivalently,
the quadratic function $q_{1}^{2}+q_{2}^{2}+p_{1}^{2}+p_{2}^{2}$. The
momentum map associated with the symplectic action of $U(2)$ on $\mathcal{R}%
^{4}$ is given by
\[
\hat{F}:\mathcal{H}\rightarrow u^{\ast }(n),
\]
\[
\left(
\begin{array}{c}
z_{1} \\
z_{2}
\end{array}
\right) \rightarrow -i\left(
\begin{array}{clcr}
z_{1}z_{1}^{\ast } & z_{1}z_{2}^{\ast } \\ z_{2}z_{1}^{\ast } &
z_{2}z_{2}^{\ast }
\end{array}
\right) .
\]
The multiplication by the imaginary unit $i$ turns it into the infinitesimal
generator of one-parameter group of unitary transformations.

Fundamental tensors for this example are given by
\begin{eqnarray*}
&&1)\quad \Delta =p_{1}\frac{\partial }{\partial p_{1}}+p_{2}\frac{\partial
}{\partial p_{2}}+q_{1}\frac{\partial }{\partial q_{1}}+q_{2}\frac{\partial
}{\partial q_{2}}\,, \\
&&2)\quad J=dp_{1}\otimes \frac{\partial }{\partial q_{1}}-dq_{1}\otimes
\frac{\partial }{\partial p_{1}}+dp_{2}\otimes \frac{\partial }{\partial
q_{2}}-dq_{2}\otimes \frac{\partial }{\partial p_{2}}\,, \\
&&3)\quad J(\Delta )=p_{1}\frac{\partial }{\partial q_{1}}-q_{1}\frac{%
\partial }{\partial p_{1}}+p_{2}\frac{\partial }{\partial q_{2}}-q_{2}\frac{%
\partial }{\partial p_{2}}\,.
\end{eqnarray*}
Therefore, to have $\hat{F}$ equivariant with respect to the infinitesimal
action of $\Delta $ and $J(\Delta )$, we have to redefine the normalized
momentum map
\[
\tilde{F}:\left(
\begin{array}{c}
z_{1} \\
z_{2}
\end{array}
\right) \rightarrow -i\rho _{z}=\frac{-i}{z_{1}z_{1}^{\ast
}+z_{2}z_{2}^{\ast }}\left(
\begin{array}{clcr}
z_{1}z_{1}^{\ast } & z_{1}z_{2}^{\ast } \\ z_{2}z_{1}^{\ast } &
z_{2}z_{2}^{\ast }
\end{array}
\right)
\]
with $i\rho _{z}\in u(n)$.

In terms of Pauli matrices, we find
\[
i\rho_z=\frac{i}{2}\left(\sigma_0+\vec x\vec\sigma\right)
\]
with identification
\[
x_1=\frac{z_1z_2^*+z_1^*z_2}{z_1z_1^*+z_2z_2^*}\,, \quad x_2=i\,\frac{%
z_1z_2^*-z_1^*z_2}{z_1z_1^*+z_2z_2^*}\,, \quad x_3=\frac{z_1z_1^*-z_2z_2^*}{%
z_1z_1^*+z_2z_2^*}
\]
and, obviously, $x_1^2+x_2^2+x_3^2=1.$

It is quite clear that the pull-back of these functions to $\mathcal{R}%
^4-\{0\}$ are invariant under the infinitesimal action of $\Delta$ and under
the infinitesimal action of $J(\Delta)$.

The Hermitian tensor introduced in sect. 3 has the form
\[
(ds)^{2}=\frac{dz_{1}^{\ast }dz_{1}+dz_{2}^{\ast }dz_{2}}{z_{1}z_{1}^{\ast
}+z_{2}z_{2}^{\ast }}-\frac{(z_{1}^{\ast }dz_{1}+z_{2}^{\ast
}dz_{2})(z_{1}dz_{1}^{\ast }+z_{2}dz_{2}^{\ast })}{(z_{1}z_{1}^{\ast
}+z_{2}z_{2}^{\ast })^{2}}
\]
providing us with the Riemannian (real part) and symplectic two-form
(imaginary part), on the complex projective metric space.

It is quite instructive to use real coordinates to write real and imaginary
part of the previous Hermitian tensor.

We set, $a\in (1,2)$
\[
z_{a}=q_{a}+ip_{a},\qquad H_{a}=\frac{1}{2}\left( p_{a}^{2}+q_{a}^{2}\right)
,\qquad d\varphi _{a}=\frac{q_{a}dp_{a}-p_{a}dq_{a}}{2H_{a}}
\]
moreover, $H=H_{1}+H_{2}$.

We find
\begin{eqnarray*}
g_{\mathrm{FS}} &=&\sum_{a}\left( \frac{dq_{a}\otimes dq_{a}+dp_{a}\otimes
dp_{a}}{2H}-\frac{dH\otimes dH}{(2H)^{2}}-\frac{4(H_{a}d\varphi _{a})\otimes
(H_{a}d\varphi _{a})}{(2H)^{2}}\right) \,, \\
\omega _{\mathrm{FS}} &=&\sum_{a}\frac{dq_{a}\wedge dp_{a}}{2H}-\frac{%
dH\wedge (2H_{1}d\varphi _{1}+2H_{2}d\varphi _{2})}{(2H)^{2}} \\
&=&\sum_{a}\frac{1}{2}\,d\left( \frac{H_{a}d\varphi _{a}}{H}\right) =\frac{1%
}{2}\sum_{a}\left( \frac{dH_{a}\wedge d\varphi _{a}}{H}\right) -\frac{%
dH\wedge (H_{1}d\varphi _{1}+H_{2}d\varphi _{2})}{(H)^{2}}
\end{eqnarray*}
By computation we find that indeed vertical vector fields are in the kernel
of the symmetric tensor and of the skew-symmetric one:
\[
g_{\mathrm{FS}}(\Delta ,\Delta )=0,\qquad g_{\mathrm{FS}}\Big(J(\Delta
),J(\Delta )\Big)=0,
\]
\[
\omega _{\mathrm{FS}}(\Delta )=0,\qquad \omega _{\mathrm{FS}}\Big(J(\Delta )%
\Big)=0.
\]
In this example, we see very clearly that $g_{\mathrm{FS}}$ is only
conformally related to the Euclidean product evaluated on horizontal vectors.

The projection (momentum map) relates the Poisson bracket on $\mathcal{R}%
^{3}\supset S^{2}$ with the Poisson brackets on $\mathcal{R}^{4}$, it is a
symplectic realization of the Poisson brackets on $\mathcal{R}^{3}$. By
considering convex combinations, we get the unit ball out of the sphere $%
S^{2}$, we have $\rho =\sum_{k}p_{k}\rho _{k}$. The space $\mathcal{D}%
\mathcal{(}C^{2})$ would be represented by density states $\frac{1}{2}\left(
\sigma _{0}+\vec{x}\vec{\sigma}\right) $ with $||\vec{x}||\leq 1$. The
topological boundary $\partial \mathcal{D}(\mathcal{C}^{2})=\mathcal{C}%
\mathcal{P}^{1}$, however, in higher dimensions it is not true that pure
states coincide with the topological boundary of density states.

The Poisson bracket extended to $\mathcal{D}(\mathcal{C}^{2})\subset
\mathcal{R}^{3}$, i.e., to functions $f_{A}(\rho )=\mbox{Tr}\,\rho A$ for
all Hermitian operators $A$, gives rise to the natural Poisson bracket on
the dual of the Lie algebra. In the present case, it is the one associated
with $SU(2)$, namely,
\[
\widetilde{\Lambda }=x_{1}\frac{\partial }{\partial x_{2}}\wedge \frac{%
\partial }{\partial x_{3}}+x_{2}\frac{\partial }{\partial x_{3}}\wedge \frac{%
\partial }{\partial x_{1}}+x_{3}\frac{\partial }{\partial x_{1}}\wedge \frac{%
\partial }{\partial x_{2}}
\]
while the metric tensor on $su^{\ast }(2)$ is
\[
\widetilde{G}=\frac{\partial }{\partial x_{1}}\otimes \frac{\partial }{%
\partial x_{1}}+\frac{\partial }{\partial x_{2}}\otimes \frac{\partial }{%
\partial x_{2}}+\frac{\partial }{\partial x_{3}}\otimes \frac{\partial }{%
\partial x_{3}}\,.
\]
The resulting partial complex structure has the form
\begin{eqnarray*}
J &=&\frac{\partial }{\partial x_{1}}\otimes \frac{x_{2}dx_{3}-x_{3}dx_{2}}{%
\sqrt{x_{2}^{2}+x_{3}^{2}}}+\frac{\partial }{\partial x_{2}}\otimes \frac{%
x_{3}dx_{1}-x_{1}dx_{3}}{\sqrt{x_{1}^{2}+x_{3}^{2}}}+\frac{\partial }{%
\partial x_{3}}\otimes \frac{x_{1}dx_{2}-x_{2}dx_{1}}{\sqrt{%
x_{1}^{2}+x_{2}^{2}}} \\
&=&\left( x_{3}\frac{\partial }{\partial x_{2}}-x_{2}\frac{\partial }{%
\partial x_{3}}\right) \otimes \frac{dx_{1}}{\sqrt{x_{2}^{2}+x_{3}^{2}}}%
+\left( x_{2}\frac{\partial }{\partial x_{1}}-x_{1}\frac{\partial }{\partial
x_{2}}\right) \otimes \frac{dx_{3}}{\sqrt{x_{1}^{2}+x_{2}^{2}}} \\
&&+\left( x_{1}\frac{\partial }{\partial x_{3}}-x_{3}\frac{\partial }{%
\partial x_{1}}\right) \otimes \frac{dx_{2}}{\sqrt{x_{1}^{2}+x_{3}^{2}}}
\end{eqnarray*}
and moreover
\[
J(x_{1}dx_{1}+x_{2}dx_{2}+x_{3}dx_{3})=0,\qquad J\left( x_{1}\,\frac{%
\partial }{\partial x_{1}}+x_{2}\,\frac{\partial }{\partial x_{2}}+x_{3}\,%
\frac{\partial }{\partial x_{3}}\right) =0.
\]
We notice that a two-form which provides a left inverse for $\Lambda $ is
given by
\[
\omega =\frac{1}{x_{1}^{2}+x_{2}^{2}+x_{3}^{2}}\left( x_{1}dx_{2}\wedge
dx_{3}+x_{2}dx_{3}\wedge dx_{1}+x_{3}dx_{1}\wedge dx_{2}\right)
\]
showing that $\omega $ is not closed! Indeed it should be closed only on
each symplectic orbit.

The ''quadratic'' function associated with $A=\left(
\begin{array}{cc}
a_{1} & a_{2} \\
a_{3} & a_{4}
\end{array}
\right) $ is given by
\begin{eqnarray*}
f_{A}(\rho ) &=&\frac{1}{2}\,\mbox{Tr}\left( \sigma _{0}+\vec{x}\vec{\sigma}%
\right) A \\
&=&\frac{1}{2}\,x_{3}(a_{1}-a_{4})+\frac{1}{2}\,x_{1}(a_{3}+a_{2})+\frac{i}{2%
}\,x_{2}(a_{2}-a_{3})+\frac{1}{2}(a_{1}+a_{4})
\end{eqnarray*}
with corresponding Hamiltonian vector field
\[
\widetilde{\Lambda }(df_{A})=\frac{a_{1}-a_{4}}{2}\left( x_{2}\frac{\partial
}{\partial x_{1}}-x_{1}\frac{\partial }{\partial x_{3}}\right) +\frac{%
a_{3}+a_{2}}{2}\left( x_{3}\frac{\partial }{\partial x_{2}}-x_{2}\frac{%
\partial }{\partial x_{3}}\right)
\]
\[
+\frac{i(a_{2}-a_{3})}{2}\left( x_{1}\frac{\partial }{\partial x_{3}}-x_{3}%
\frac{\partial }{\partial x_{1}}\right) .
\]
By considering also the gradient vector field associated with Casimir
function $\zeta =x_{1}^{2}+x_{2}^{2}+x_{3}^{2}$, we get
\[
\widetilde{\Delta }=\tilde{G}(d\zeta ),\qquad \widetilde{\Delta }=x_{1}\frac{%
\partial }{\partial x_{1}}+x_{2}\frac{\partial }{\partial x_{2}}+x_{3}\frac{%
\partial }{\partial x_{3}}\,,
\]
which along with the three rotation vector fields provides a basis for the
module of vector fields on the unit ball.

A decomposition of a generic linear vector field on the unit ball can be
achieved by using the basis $\{x_j\frac{\partial}{\partial x_k}\}$, $%
j,k\in\{1,2,3\}$.

\section{Composite systems}

The state space of a composite system is the tensor product of the state
spaces of the component systems. If $\mathcal{H}_A$ and $\mathcal{H}_B$ are
the Hilbert spaces of the component systems, the Hilbert space for the
composite system is $\mathcal{H}= \mathcal{H}_A\otimes\mathcal{H}_B$.

Clearly, once $\mathcal{H}$ has been built, we could use all constructions
we have already performed in the previous sections. Here we would like to
keep track of the component systems and of the geometrical structures
pertaining to them. Instead of general aspects, we shall concentrate
directly on an example. We consider component systems to be two-level
quantum systems, i.e., $\mathcal{H}=\mathcal{C}^2\otimes\mathcal{C}^2\equiv%
\mathcal{C}^4$.

If $\left|
\begin{array}{c}
z_1\cr z_2
\end{array}
\right|,\quad \left|
\begin{array}{c}
w_1\cr w_2
\end{array}
\right|$ are state vectors for $\mathcal{H}_A$ and $\mathcal{H}_B$,
respectively, we have
\[
\left|
\begin{array}{c}
z_1\cr z_2
\end{array}
\right|\otimes \left|
\begin{array}{c}
w_1\cr w_2
\end{array}
\right|= \left|
\begin{array}{c}
u_1\cr u_2\cr u_3\cr u_4
\end{array}
\right|=\left|
\begin{array}{c}
z_1w_1\cr z_1w_2\cr z_2w_1\cr z_2w_2
\end{array}
\right|.
\]
The momentum map, which imbeds the complex projective space of the composite
system into the Lie algebra of $U(4)$ is given by
\[
\rho_u=\left|
\begin{array}{c}
u_1\cr u_2\cr u_3\cr u_4
\end{array}
\right|
\begin{array}{clcr}
|u_1^* & u_2^* & u_3^* & u_4^*|\cr ~ & ~ & ~ & ~\cr ~ & ~ & ~ & ~\cr~ & ~ & ~
& ~
\end{array}
= \left|
\begin{array}{clcr}
u_1u_1^* & u_1u_2^* & u_1u_3^* & u_1u_4^*\cr u_2u_1^* & u_2u_2^* & u_2u_3^*
& u_2u_4^*\cr u_3u_1^* & u_3u_2^* & u_3u_3^* & u_3u_4^*\cr u_4u_1^* &
u_4u_2^* & u_4u_3^* & u_4u_4^*
\end{array}
\right|.
\]
Using the representation of the density states for the component systems in
terms of Pauli matrices, we find pure separable states for the composite
system described by
\[
\rho=\frac{1}{4}\Big(\mathbf{1}+n_j\sigma_A^j\otimes\mathbf{1}_B +m_k\mathbf{%
1}_A\otimes\sigma_B^k+n_jm_k\sigma_A^j\otimes\sigma_B^k \Big)
\]
with $\|\vec n\|^2=\|\vec m\|^2=1$.

In general, a density state will have the form
\[
\rho =\frac{1}{4}\Big(\mathbf{1}+p_{j}\sigma _{A}^{j}\otimes \mathbf{1}%
_{B}+q_{k}\mathbf{1}_{A}\otimes \sigma _{B}^{k}+r_{jk}\sigma _{A}^{j}\otimes
\sigma _{B}^{k}\Big)
\]
with the condition $\sum_{j}(p_{j}^{2}+q_{j}^{2})+\sum_{j,k}r_{jk}^{2}\leq 1$%
.

Matrices $i(\sigma _{A}^{j}\otimes \mathbf{1}_{B})$, $i\mathbf{1}_{A}\otimes
\sigma _{B}^{k})$, $i(\sigma _{A}^{j}\otimes \sigma _{B}^{k})$ are a basis
for the Lie algebra of $SU(4)$, thus adding to them the identity matrix $i%
\mathbf{1}=i(\mathbf{1}_{A}\otimes \mathbf{1}_{B})$, we get a basis for the
Lie algebra $u(4)$. Any Hermitian matrix, after multiplication by the
imaginary unit $i$, can be decomposed in previous basis. In particular, any $%
\rho_{u}$ can be rewritten by means of previous basis.

In terms of a basis for a Cartan subalgebra of $iu(n)$, say,
\[
\lambda_0=\left|
\begin{array}{clcr}
1 & ~ & ~ & ~\cr ~ & 1 & ~ & ~\cr~ & ~ & 1 & ~\cr~ & ~ & ~ & ~1
\end{array}
\right|,\quad \lambda_2=\left|
\begin{array}{clcr}
1 & ~ & ~ & ~\cr ~ & 1 & ~ & ~\cr~ & ~ & -1 & ~\cr~ & ~ & ~ & ~-1
\end{array}
\right|,
\]
\[
\lambda_1=\left|
\begin{array}{clcr}
1 & ~ & ~ & ~\cr ~ & -1 & ~ & ~\cr~ & ~ & 1 & ~\cr~ & ~ & ~ & ~-1
\end{array}
\right|,\quad \lambda_3=\left|
\begin{array}{clcr}
1 & ~ & ~ & ~\cr ~ & -1 & ~ & ~\cr~ & ~ & -1 & ~\cr~ & ~ & ~ & ~1
\end{array}
\right|
\]
corresponding to $\lambda_0=\mathbf{1}\otimes\mathbf{1}$, $%
\lambda_1=\sigma_3\otimes\mathbf{1}$, $\lambda_2=\mathbf{1}\otimes\sigma_3$,
$\lambda_3=\sigma_3\otimes\sigma_3$.

We may write a generic density state in the form
\[
U\frac{1}{4}\left(\lambda_0+p^1\lambda_1+p^2\lambda_2+
p^3\lambda_3\right)U^\dagger=\frac{1}{4}\left(\lambda_0+\vec p
\vec\lambda\right),
\]
where $\vec p$ is a vector in the 15-dimensional space and $\vec\lambda$
stays for a "vector of matrices" in the 15-dimensional Lie algebra $su(4)$.

In terms of the states of the component systems, we can now express
operators as "quadratic functions" and compute the Riemann--Jordan bracket
and the Poisson bracket in terms of those of the component systems.

By using evident notation, we may consider "quadratic functions"
\[
\langle\psi\mid\otimes\langle\varphi\mid A\otimes B\mid\varphi\rangle
\otimes\mid\psi\rangle=\langle\psi\mid A\mid\psi\rangle\otimes
\langle\varphi\mid B\mid\varphi\rangle,
\]
which are really "biquadratic" if we parametrize states in terms of the
states of the component systems.

The Poisson bracket is defined by
\[
\{f_A\otimes f_B,g_A\otimes g_B\}=\{f_A,g_A\}\otimes
f_Bg_B+f_Ag_A\otimes\{f_B,g_B\},
\]
more specifically
\[
\{z_mw_n,z_rw_s\}=\{z_m,z_r\} w_nw_s+z_mz_r\{w_n,w_s\}
\]
and similarly for the Riemann--Jordan brackets.

\section{Conclusions and outlook}

In this paper, we have shown how it is possible to provide a geometrical
formulation of quantum mechanics in a way that makes possible to use also
nonlinear transformations. Not only it seems possible to achieve a great
level of geometrization of quantum mechanics comparable to the one obtained
in classical mechanics and general relativity but, in addition, elsewhere we
will show how to put to work this covariant formulation of quantum mechanics
for a full study of composite systems and how to tackle the problem of
separability and entanglement.

\section*{Acknowledgments}

V.~I.~M. and E.~C.~G.~S. thank Dipartimento di Scienze Fisiche,
Universit\'{a} ``Federico~II'' di Napoli and Istituto Nazionale di
Fisica Nucleare, Sezione di Napoli for kind hospitality.


\begin{thebibliography}{99}

\bibitem{Dirac}
P.A.M. Dirac, The Principles of Quantum Mechanics, 4th edition
(Pergamon, Oxford, 1958)

\bibitem{Carinena}  J.F. Cari\~nena, J. Grabowski, G. Marmo, Lee--Scheffers
Systems: A Geometrical Approach (Bibliopolis, Napoli,
2000)\newline J.F. Cari\~nena, G. Marmo, J. Nasarre, The nonlinear
superposition principle and the Weyl--Norman method, Int. J. Mod.
Phys. \textbf{A 13} (1998) 3601--362

\bibitem{Schroedinger}  E. Schr\"{o}dinger, Zum Heisenbergshen
Unsh\"{a}rfeprinzip, Ber. Kgl. Akad. Wiss. \textbf{296} (1930) 296--303

\bibitem{MMSZ}  V. I. Man'ko, G. Marmo, E. C. G. Sudarshan and F. Zaccaria,
Interference and entanglement: an intristic approach, J. Phys. A:
Math. Gen. \textbf{35} (2002) 7173--7157; Inner composition law of
pure states, Phys. Lett. \textbf{A 273} (2000) 31--; Inner
composite law of pure spin states, pp. 92--97 in Spin-Statistics
Connection and Commutation Relations, R. C. Hilborn and G. M. Tino
(Eds.) American Institute of Physics, Melville, N.Y. 2000, Vol.
545

\bibitem{Gorelli}  R. Cirelli, M. Gatti, A. Mani\'{a}, On the nonlinear
extension of quantum superposition and uncertainty principles, J. Geometry
and Physics \textbf{29} (1999) 54--86

\bibitem{Jordan}  P. Jordan, J. von Neumann, E. Wigner, On an algebraic
generalization of the quantum mechanical formalism, Ann. Math. \textbf{35}
(1934) 29--54

\bibitem{all}  R. Cirelli, P. Lanzavecchia, A. Mani\'{a}, Normal pure states
of the von Neumann algebra of bounded operators as K\"{a}hler
manifold, J. Phys. A: Math. Gen. \textbf{15} (1983)
3829--3835\newline R. Cirelli, P. Lanzavecchia, Hamiltonian vector
fields in quantum mechanics, Il. Nuovo Cimento \textbf{B 79}
(1984) 271--283\newline M. C. Abbati, R. Cirelli, P. Lanzavecchia,
A. Mani\'{a}, Pure states of general quantum-mechanical systems as
K\"{a}hler Bundle, Il. Nuovo Cimento \textbf{B 83} (1984)
43--60\newline A. Bloch, An infinite-dimensional Hamiltonian
system on a projective Hilbert
space, Transactions of the Am. Math. Soc. \textbf{302} (1987) 787--796%
\newline
V. Cantoni, The Riemannian structure on the space of quantum-like systems,
Comm. Math. Phys. \textbf{55} (1977) 189--193; Intrinsic geometry of the
quantum-mechanical phase space, Hamiltonian systems and correspondence
principle, Licei. Rend. Sc. Fis. Mat. e Nat. \textbf{LXII} (1977) 628--636%
\newline
A. Heslot, Quantum mechanics as a classical theory, Phys. Rev.
\textbf{D 31} (1985) 1341--1348\newline D. J. Rowe, A. Ryman, G.
Rosensteel, Many-body quantum mechanics as a
symplectic dynamical system, Phys. Rev. \textbf{A 22} (1980) 2362--2373%
\newline
T. R. Field, J. S. Anandan, Geometric phases and coherent states, J. Geom.
Phys. \textbf{50} (2004) 56--78\newline
D. C. Brody, L. P. Hughston, Geometric quantum mechanics, J. Geom. Phys.
\textbf{38} (2001) 19--53\newline
F. Strocchi, Complex coordinates and quantum mechanics, Rev. Mod. Phys.
\textbf{38} (1956) 36--40\newline
A. Ashtekar, T. A. Shilling, Geometrical formulation of quantum mechanics,
gr-qc/9706059, in On Einstein's Path, A. Harvey (Ed.) Springer-Verlag,
Berlin (1998)

\bibitem{Ch-Jam}  D. Chruscinski, A. Jamiolkowski, Geometric Phases in
Classical and Quantum Mechanics (2004) Birkh\"{a}user, Boston

\bibitem{Pizzocdeura}  R. Cirelli, A. Mani\'{a}, L. Pizzocchero,Quantum
mechanics as an infinite-dimensional Hamiltonian system with
uncertainty structure. I. J. Math. Phys. \textbf{31} (1990)
2891--2897; II. J. Math. Phys. \textbf{31} (1990) 2898--2903

\bibitem{EMS}  G. Esposito, G. Marmo, G. Sudarshan, From Classical to
Quantum Mechanics (2004) Cambridge University Press, Cambridge

\bibitem{Russ}  V.I. Man'ko, G. Marmo, E.C.G. Sudarshan, F. Zaccaria,
On the relation between Schroedinger and von Neumann equation, J.
Russ. Laser Res. \textbf{20} (5) (1999) 421-437

\bibitem{Grab}  J. Grabowski, G. Landi, G. Marmo, G. Vilasi, Generalized
reduction procedure, Fortschr. Phys. \textbf{42}(5) (1994) 393-427

\bibitem{Study}  E. Study, Kuerzeste Wege in Komplexen Gebiet, Math. Annalen
\textbf{60} (1905) 321; G. Fubini, Sulle metriche definite da una
forma Hermitiana, Atti Istituto Veneto \textbf{6} (1903) 501

\bibitem{Aldaya}  V. Aldaya, J. Guerrero, G. Marmo, Quantization on a Lie
group: Higher-order polarizations, in Symmeries in Science X, B. Gruber and
M. Ramek (Eds) Plenum Press, New York (1998) pp. 1--35

\bibitem{Mukunda}  E. Ercolessi, G. Marmo, G. Morandi, N. Mukunda, Geometry
of mixed states and degeneracy structure of geometric phases for
multi-level quantum systems. A unitary group approach, Int. J.
Mod. Phys. \textbf{A 15} (2001) 5007--5032

\bibitem{Marmo}  D. Alekseevsky, J. Grabowski, G. Marmo, P. W. Michor,
Poisson structures on double Lie groups, J. Geom. Phys. \textbf{26} (1998)
340--379

\bibitem{Benvegnu}  A. Benvegn\'{u}, N. Sansonetto, M. Spera, Remarks on
Geometrical Quantum Mechanics, J. Geom. Phys. \textbf{51} (2004)
229--243

\end{thebibliography}
\end{document}